\documentclass[preprint, amsmath,amssymb, nofootinbib]{revtex4}

\usepackage[]{caption}
\usepackage{color, graphicx, bm}

\usepackage{float}

\newcommand{\bea}{\begin{eqnarray}}
\newcommand{\ena}{\end{eqnarray}}

\renewcommand{\a}{\alpha}

\renewcommand{\d}{\delta}

\newcommand{\e}{\epsilon}

\renewcommand{\l}{\lambda}

\begin{document}

\title{\Large\bf Observational constraints on slow-roll inflation coupled to
a Gauss-Bonnet term}

\author{Seoktae Koh} \email[email: ]{kundol.koh@jejunu.ac.kr}

\affiliation{
Department of Science Education, Jeju National University,
 Jeju, 690-756, Korea }

\author{Bum-Hoon Lee} \email[email: ]{bhl@sogang.ac.kr}

\author{Wonwoo Lee} \email[email: ]{warrior@sogang.ac.kr}

\author{Gansukh Tumurtushaa} \email[email: ]{gansuh@sogang.ac.kr}

\affiliation{
 Center for Quantum Spacetime, Sogang University, Seoul 121-742, Korea \\
 Department of Physics, Sogang University, Seoul 121-742, Korea}


\begin{abstract}
We study slow-roll inflation with a Gauss-Bonnet term that is coupled to
an inflaton field nonminimally. We investigate the inflationary solutions for a specific type of the nonminimal coupling to the Gauss-Bonnet term and  inflaton potential both analytically and numerically. We also calculate the observable quantities such as the
power spectra of the scalar and tensor modes, the spectral indices, the tensor-to-scalar ratio
and the running spectral indices.
 Finally, we constrain our result with the observational data by Planck and BICEP2 experiment.
\end{abstract}

\maketitle



\section{Introduction} \label{intro}

Recent experiments and observations including Planck \cite{Ade:2013zuv},
LHC \cite{Chatrchyan:2012ufa}, and
BICEP2 \cite{Ade:2014xna} confirmed that the inflation paradigm is believed to be successful
for explaining the evolution of our Universe and generation
of large scale structure formation. The cosmic microwave background (CMB) observations by Planck and WMAP imply that our Universe is Gaussian, adiabatic, and nearly scale invariant.
Although there are some debates \cite{Ijjas:2013vea},
the Planck data seem to favor the  inflationary model with the
simple scalar field potential, especially the convex-type potential
\cite{Ade:2013uln}. But recent BICEP2 combined with Planck data
seems to favor the concave-type potential, especially
the $\phi^2$ potential.

While Planck and  WMAP provide the upper bound on the tensor-to-scalar ratio ($ r < 0.12$), recent BICEP2 telescope \cite{Ade:2014xna} at the South Pole reported the detection of B-mode polarization signal \cite{Michael:2014},\cite{Raphael:2014} which is generated by the tensor perturbation (gravitational wave modes) in an inflationary period.
According to BICEP2, $r = 0.20$ at $5.2 \sigma$
with $r=0$ disfavored  at $7.0 \sigma$. This tensor-to-scalar ratio
value is larger than the upper bound by  Planck + WMAP.
It has been widely studied how to reconcile this
discrepancy between two data, and one simple resolution, which was suggested
in Ref. \cite{Ade:2014xna}, is to consider  the running spectral index,
$d n_s/d\ln k$.

Although inflation is believed to solve a lot of the outstanding problems
of the standard big bang cosmology such as the horizon and flatness problem,
there are still several unsolved problems in an inflation scenario, for example, the flat potential problem, initial singularity problem, and  quantum gravity (trans-Planckian problem).

Especially, if we think over the very early Universe approaching the Planck scale,
we could consider Einstein gravity with some corrections
as the effective theory of the ultimate quantum gravity.
For instance, the higher derivative  terms of gravity
with nontrivial gravitational self-interactions naturally appear
in the low energy limits of string theories.
The presence of curvature squared terms such as a Gauss-Bonnet (GB)
combination does not have any ghost particles as well as
any problem with the unitarity.
Additionally, the order of the gravitational equation of motion,
the second-order derivatives of the metric tensor,
does not change if there is no nonminimal coupling to a Gauss-Bonnet term \cite{Callan:1985ia}.
Fortunately, the theory with a nonminimally coupled Gauss-Bonnet term
could provide the possibility of avoiding the initial
singularity of the Universe \cite{Antoniadis:1992rq}.
It may violate the energy condition thanks to the presence of the
term in the singularity theorem \cite{Hawking:1969sw}.
In this perspective, one could introduce the Einstein theory
of gravity having a scalar field with a nonminimally coupled
Gauss-Bonnet term as the effective theory added a quantum correction.

Generally, the Gauss-Bonnet term in four dimensions is known as  the topological term,
so the dynamics is not influenced by the Gauss-Bonnet term.
In order to consider the effect of the Gauss-Bonnet term on the spacetime as well as
the field evolution, the Gauss-Bonnet term is required to be coupled to the matter field.
Recently a number of papers with this motivation were studied \cite{Hwang:2000},\cite{Kawai:1999} and discussed phenomenology in detail in \cite{Satoh:2008},\cite{Satoh:2008a},\cite{Satoh:2010},\cite{Guo:2010jr} and \cite{Jiang:2013gza}.
In Refs. \cite{Guo:2010jr},\cite{Jiang:2013gza},
the authors studied the specific inflationary model with the  Gauss-Bonnet term
constrained by the  WMAP data in \cite{Guo:2010jr} and
by the  Planck data in \cite{Jiang:2013gza}.
They analytically derived the power spectra of the scalar
and tensor perturbations. They employed a monomial potential
and an inverse monomial Gauss-Bonnet coupling that satisfies $V(\phi) \xi(\phi) \sim  \text{const}$.
These choices of the  potential and  Gauss-Bonnet coupling
provide the relatively large parameter values, $\alpha
\equiv 4\kappa^4 V_0 \xi_0/3 \sim \mathcal{O}(10^{-2})$,
to be consistent with observations and
 showed that a positive (or negative)
coupling leads to a reduction (or enhancement)
of the tensor-to-scalar ratio.

In this work, we try to relax the condition $V(\phi) \xi(\phi) \sim \text{const}$
and then constrain from the recent observations by Planck and BICEP2.
We also calculate the spectral indices of the scalar and tensor perturbations,
 the tensor-to-scalar ratio and the running spectral  index.

The outline of this paper is as follows: In Sec. \ref{sect2},
 we set up the basic framework with the Gauss-Bonnet term for this work.
The $e$-folding numbers are calculated and then give a constraint on
the model parameter.
In Sec. \ref{sect3}, we briefly review the linear perturbations
with the Gauss-Bonnet coupling term
 and then calculate the observable quantities
 such as the power spectra, the spectral indices, the tensor-to-scalar ratio
and the running spectral indices.
In Sec.\ \ref{sect4}, we examine the specific models consistent
 with our motivations. We compare our result
 with the observational data by the Planck  data and recent BICEP2.
Finally, we summarize our results in Sec. \ref{summary}.


\section{Slow-roll inflation with the GB term} \label{sect2}

We consider  an  action with the Gauss-Bonnet term that is coupled to a scalar field
\begin{align}
S = &\int_{\mathcal{M}} d^4x\sqrt{-g}
\left[\frac{1}{2\kappa^2} R - \frac{1}{2}g^{\mu\nu}
\partial_{\mu}\phi \partial_{\nu} \phi
 - V(\phi)-\frac12\xi(\phi) R_{\rm GB}^2\right],
\label{action}
\end{align}
where $\phi$ is an inflaton field with a potential
$V(\phi)$, $R$ is the Ricci scalar curvature of the spacetime $\mathcal{M}$, $R^{2}_{\rm GB}
= R_{\mu\nu\rho\sigma} R^{\mu\nu\rho\sigma} - 4 R_{\mu\nu}
R^{\mu\nu} + R^2$ is the Gauss-Bonnet  term, and $\kappa^2 = 8\pi G$.
 The Gauss-Bonnet coupling $\xi(\phi)$ is required to be
 a function of a scalar field in order to give nontrivial effects
on the background dynamics.

 Varying the action (\ref{action}) with respect to $g_{\mu\nu}$ and
$\phi$ yields the Einstein and field equation
\begin{align}
& R_{\mu\nu} - \frac{1}{2}g_{\mu\nu}R = \kappa^2 \left( \partial_{\mu}
\phi \partial_{\nu}\phi -\frac{1}{2}g_{\mu\nu} (g^{\rho\sigma}
\partial_{\rho} \phi \partial_{\sigma} \phi + 2 V) + T_{\mu\nu}^{GB}
\right),
\label{einstein}
\\
& \square \phi  - V_{\phi} -\frac{1}{2}T^{GB} = 0,
\label{kg}
\end{align}
where $\square \equiv \partial_{\mu} \partial^{\mu}$.
$T_{\mu\nu}^{GB}$ and $T^{GB}$ are   the energy-momentum tensor and
its trace for the  Gauss-Bonnet term, respectively,
 which are given by
\begin{align}
T_{\mu\nu}^{GB} =& 4 (\partial^{\rho}\partial^{\sigma} \xi R_{\mu\rho\nu\sigma}
-\square \xi R_{\mu\nu} + 2\partial_{\rho} \partial_{(\mu} \xi {R^{\rho}}_{\nu)}
-\frac{1}{2} \partial_{\mu}\partial_{\nu} \xi R)
\nonumber \\
&  -2(2\partial_{\rho}\partial_{\sigma} \xi
R^{\rho\sigma} - \square \xi R) g_{\mu\nu},
\\
T^{GB} =& \xi_{\phi} R_{GB}^2.
\end{align}

In a spatially flat Friedmann-Robertson-Walker universe with a scale factor
$a$,
\begin{align}
ds^2 = -dt^2 + a(t)^2 \delta_{ij} dx^i dx^j,
\end{align}
the background  Einstein and field equations yield
\bea
\label{beq2}
&& H^2 = \frac{\kappa^2}{3} \left(\frac{1}{2}\dot{\phi}^2
+ V + 12\dot{\xi}H^3 \right), \\
\label{beq3} && \dot{H} = -\frac{\kappa^2}{2}
\left( \dot{\phi}^2 - 4\ddot{\xi}H^2 - 4\dot{\xi} H
(2\dot{H} - H^2) \right), \\
&& \ddot{\phi} + 3 H \dot{\phi} + V_{,\phi} +12 \xi_{,\phi} H^2 \left(\dot{H}+H^2\right) = 0,
\label{beq4}
\ena
where a dot represents a derivative with respect to the cosmic time $t$,
$H \equiv \dot{a}/a$ denotes the Hubble parameter, and $V_{,\phi} = \partial V/\partial \phi,
\,\, \xi_{,\phi} = \partial \xi/\partial \phi$. Since $\xi$ is a function of $\phi$,
$\dot{\xi}$ implies $\dot{\xi} = \xi_{,\phi} \dot{\phi}$. If $\xi$ is a constant,
then Eqs. (\ref{beq2})--(\ref{beq4}) are reduced to those for  standard inflation
without the Gauss-Bonnet coupling.

In this work, we consider slow-roll inflation with the inflaton potential
and  Gauss-Bonnet coupling satisfying the slow-roll approximations
\begin{align}
\dot{\phi}^2/2  \ll V, \quad \ddot{\phi} \ll 3H\dot{\phi},
\quad 4\dot{\xi}H \ll 1, \quad \ddot{\xi} \ll \dot{\xi} H.
\label{sra}
\end{align}
In addition to the usual slow-roll approximations, we introduce  two more
conditions related to the Gauss-Bonnet coupling.

\begin{figure}
\centering
\includegraphics[width=1.0\textwidth]{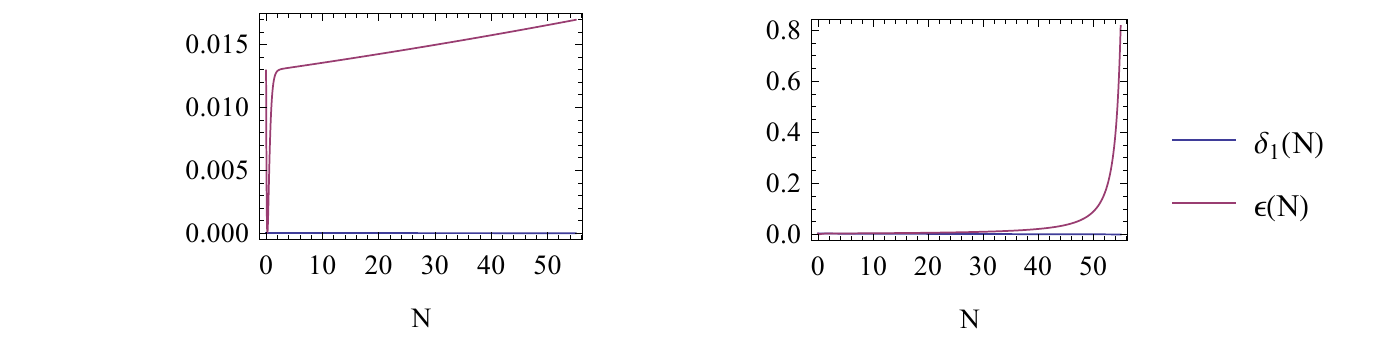}
\caption{Left: evolution of the slow-roll parameters $\epsilon$ and $\delta_1$
for $\xi(\phi) = \xi_0 e^{-\lambda \phi}$ with $\lambda =-0.1$.
\,\,Right: evolution of the slow-roll parameters $\epsilon$ and $\delta_1$
for $\xi(\phi) = \xi_0 \phi^2$.}
\label{fig_sr}
\end{figure}

To reflect these approximations, we introduce the slow-roll parameters,
\begin{align}
\epsilon &= -\frac{\dot{H}}{H^2}  , \quad
\eta = \frac{\ddot{H}}{H\dot{H}} , \quad
\zeta = \frac{\dddot{H}}{H^2 \dot{H}},
\nonumber \\
\delta_1 &= 4\kappa^2\dot{\xi}H , \quad
 \delta_2 = \frac{\ddot{\xi}}{\dot{\xi}H}, \quad
 \delta_3 = \frac{\dddot{\xi}}{\dot{\xi}H^2}.
\label{srparam}
\end{align}
We have checked the validity of the
new slow-roll parameters during an accelerating phase numerically
in Fig. \ref{fig_sr}.

If  the slow-roll approximations (\ref{sra}) are taken into account,
the background equations, (\ref{beq2})-(\ref{beq4}), reduce to
\bea
\label{seq1}
&& H^2 \simeq \frac{\kappa^2}{3}  V, \\
&& \dot{H} \simeq -\frac{\kappa^2}{2}( \dot{\phi}^2 + 4\dot{\xi}H^3), \\
&& 3H \dot{\phi} + V_{,\phi} + 12\xi_{,\phi}H^4 \simeq 0 ,
\label{seq2}
\ena
which allows us to obtain the number of $e$-folds
\bea
\label{ne}
N(\phi) = \int_{t}^{t_e} H dt \simeq \int_{\phi_{e}}^{\phi}
 \frac{3\kappa^2  V}{3 V_{,\phi}+4 \kappa^4 \xi_{,\phi}V^2}d\phi
 \equiv \int^{\phi}_{\phi_e} \frac{\kappa^2}{Q} d\phi.
\ena
where a $\phi_e$ is determined from the condition $\epsilon(\phi_e) =1$ and
\begin{align}
Q \equiv \frac{V_{,\phi}}{V} + \frac{4}{3}\kappa^4 \xi_{,\phi}  V.
\end{align}

\begin{figure}
\centering
\includegraphics[width=0.8\textwidth]{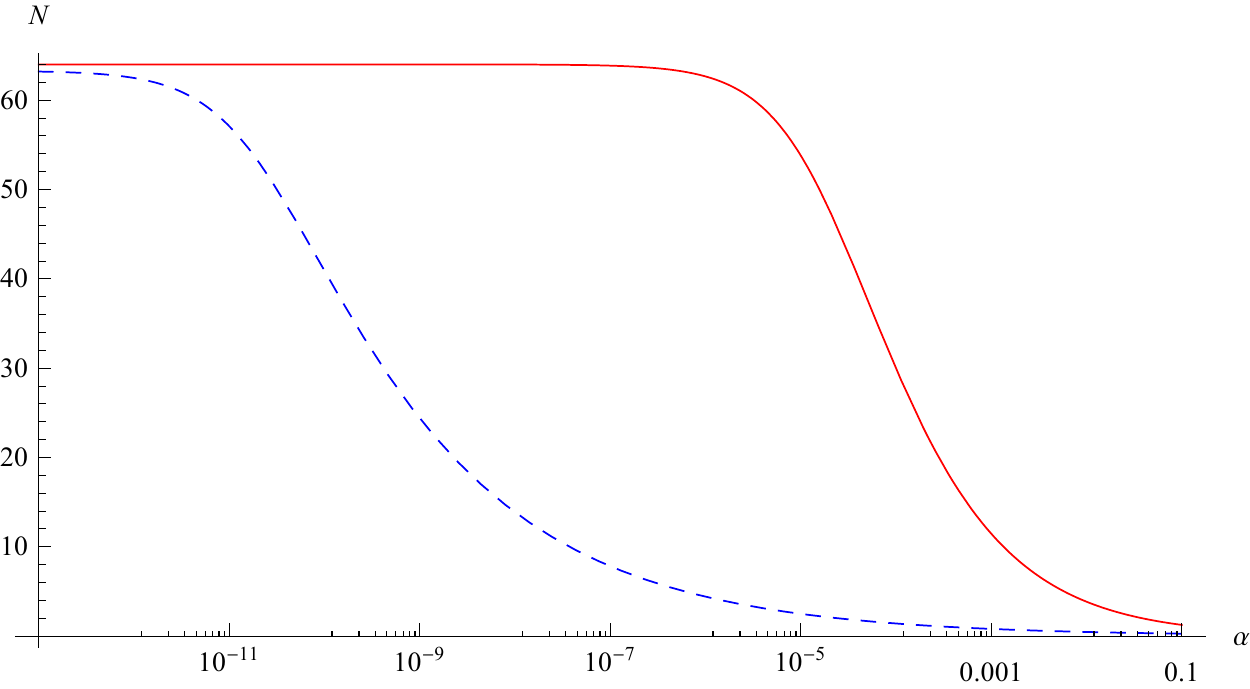}
\caption{ $e$-folding numbers with $\alpha \equiv
\frac{4}{3}\kappa^4 V_0 \xi_0$ for $V=V_0 \phi^2,\,\, \xi = \xi_0 \phi^2$
(solid line) and $V=V_0 \phi^4, \,\, \xi=\xi_0 \phi^4$ (dashed line).}
\label{fig_efolding}
\end{figure}

If we choose $V = V_0 \phi^n$ and $\xi = \xi_0 \phi^n$,
the number of $e$-folds are calculated assuming $\phi_e$ is negligible
compared to $\phi_i$ as
\begin{align}
N = \int^{\phi_i}_{\phi_e}
\frac{\kappa^2}{Q} \simeq \frac{\kappa^2 \phi_i^2}{2n} {}_2 F_1 \left(
1,\frac{1}{n},1+\frac{1}{n}; -\alpha \phi_i^{2n} \right),
\label{efold_pow}
\end{align}
where ${}_2 F_1$ is the hypergeometric function and $\alpha
\equiv \frac{4}{3}\kappa^4 V_0 \xi_0$.
We plot the number of $e$-folds $N$  with $\alpha$ for $n=2$ (solid line)
and $n=4$ (dashed line) in Fig. \ref{fig_efolding}.
The condition of $N \gtrsim 60$ requires
$\alpha \lesssim \alpha_c = 10^{-6}\,\, M_p^{-4}$ for $n=2$
and $\alpha \lesssim  \alpha_c = 10^{-12}\,\, M_p^{-8}$ for $n=4$.
Here, $\alpha_c$ is the value when $\alpha$ becomes nearly constant.
We find that $N$ approaches $N_{max}$ as $\alpha$ decreases to $\alpha_c$.
Because the hypergeometric function ${}_2 F_1$ is constant for $\alpha \lesssim
\alpha_c$, $N$ cannot become larger than $N_{max}$ unless $\phi_i$ increases.
$\phi_i = 15 M_p\,\, (n=2)$ and $\phi_i = 22 M_p\,\,(n=4)$ are
 required to obtain $N \simeq 60$ in Fig. \ref{fig_efolding}.

\begin{figure}[H]
\centering
\includegraphics[width=0.9\textwidth]{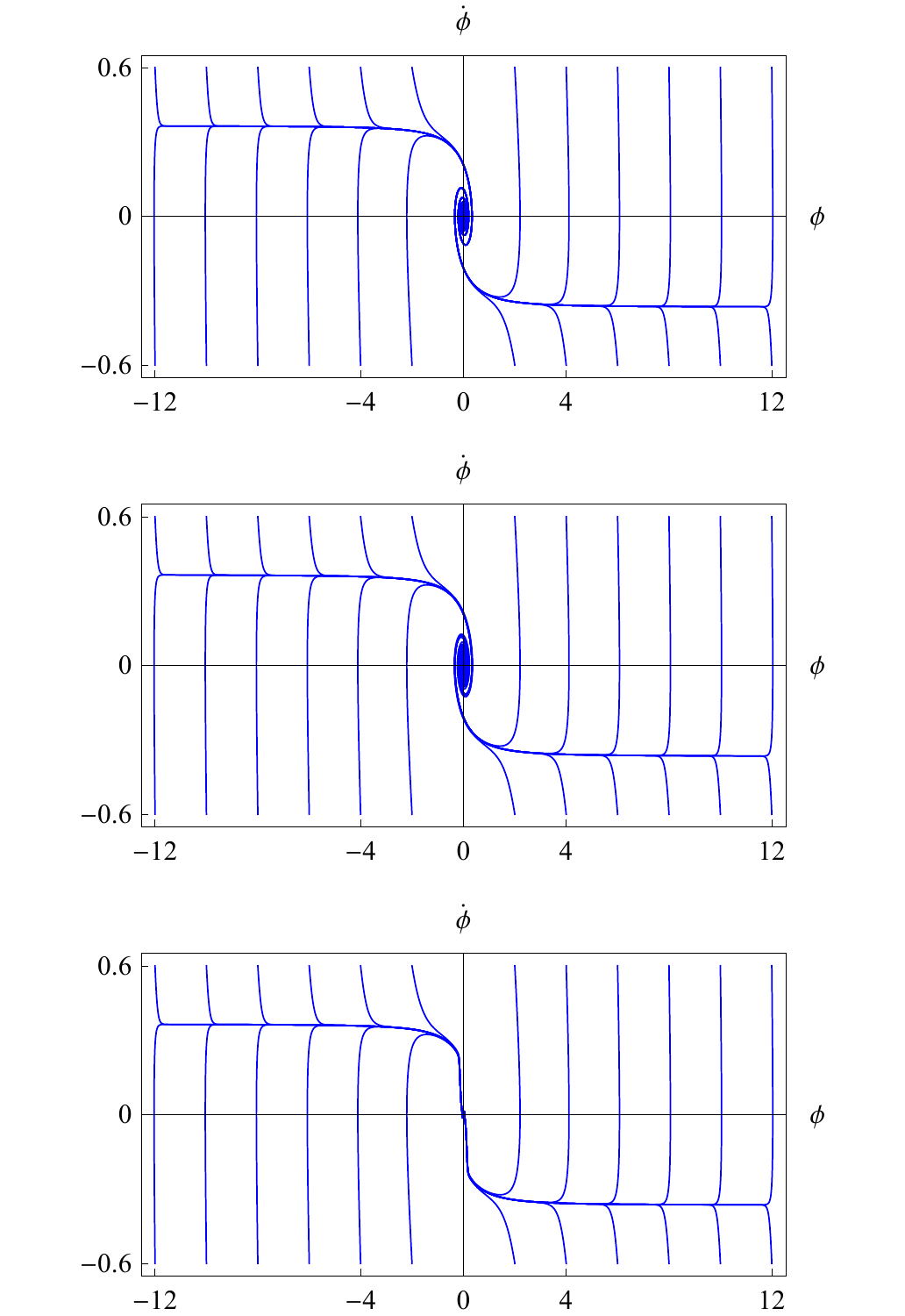}
\caption{Attractor-type solutions of the background equations of motion
for standard inflation without the Gauss-Bonnet term, chaotic inflation
with the monomial Gauss-Bonnet coupling, and chaotic inflation with the inverse
monomial Gauss-Bonnet coupling
}
\label{fig_attractor}
\end{figure}

It is convenient to express the slow-roll parameters (\ref{srparam})
in terms of the potential and  Gauss-Bonnet coupling:
\begin{align}\label{srpote}
\epsilon =& \frac{1}{2\kappa^2} \frac{V_{\phi}}{V}  Q, \\
 \eta =& -\frac{V_{\phi\phi} Q}{\kappa^2 V_{\phi}}
-  \frac{1}{\kappa^2} Q_{\phi},
\\
\zeta =& \frac{V_{\phi\phi\phi} Q^2}{\kappa^4 V_{\phi}}
+\frac{V_{\phi\phi}Q^2}{2\kappa^4 V} +\frac{3 V_{\phi\phi} Q_{\phi} Q}{\kappa^4 V_{\phi}}
+\frac{V_{\phi}Q_{\phi} Q}{2\kappa^4 V}
\nonumber \\
&  +\frac{1}{\kappa^4}Q_{\phi}^2
+\frac{1}{\kappa^4}Q_{\phi\phi}Q,
\\
\delta_1 =&  -\frac{4\kappa^2}{3} \xi_{\phi} V Q,
\\
\delta_2 =& -\frac{\xi_{\phi\phi}Q}{\kappa^2 \xi_{\phi}}
-\frac{V_{\phi}Q}{2\kappa^2 V} - \frac{1}{\kappa^2} Q_{\phi},
\\
\delta_3 =& \frac{\xi_{\phi\phi\phi} Q^2}{\kappa^4 \xi_{\phi}}
+ \frac{3\xi_{\phi\phi}V_{\phi} Q2}{2\kappa^4 \xi_{\phi}V}
+ \frac{3\xi_{\phi\phi}Q_{\phi}Q}{\kappa^4 \xi_{\phi}}
+\frac{V_{\phi\phi}Q^2}{2\kappa^4 V}
\label{srpotd}
\nonumber \\
& + \frac{2V_{\phi}Q_{\phi}Q}{\kappa^4 V}
+\frac{1}{\kappa^4}Q_{\phi}^2 + \frac{1}{\kappa^4}Q Q_{\phi\phi}.
\end{align}
Equation (\ref{seq2}) becomes for $V = V_0 \phi^2$ and $\xi = \xi_0 \phi^2$
\begin{align}
\dot{\phi} \approx -\sqrt{\frac{\alpha}{\xi_0 \kappa^6}}
(1+\alpha  \phi^4),
\label{sreq3}
\end{align}
and then we get the solution assuming $\alpha$-term is negligibly small
from Fig. \ref{fig_efolding},
\begin{align}
\phi(t) \sim -\sqrt{\frac{\alpha}{\xi_0 \kappa^6}} t + \text{const}.
\end{align}
We find that this slow-roll trajectory is the attractor solution in Fig. \ref{fig_attractor}.
The slow-roll trajectories of Eqs.(\ref{seq1})--(\ref{seq2})
were proved to be the attractor solutions generally  when the Gauss-Bonnet
term is coupled to the scalar field in Ref. \cite{Guo:2010jr}.
We compare the attractor behavior for three cases: standard chaotic
inflation (upper), chaotic inflation with the monomial Gauss-Bonnet coupling  (middle),
and chaotic inflation with the inverse monomial Gauss-Bonnet coupling (below)  (which
was considered in Ref. \cite{Guo:2010jr}) in Fig. \ref{fig_attractor}.

\section{Linear Perturbations and Power Spectra \label{sect3}}
We briefly review the linear perturbations with the Gauss-Bonnet
coupling in this section.

The linearized metric in the comoving gauge in which
$\delta \phi = 0$ takes the form
\begin{align}
ds^2 = a(\tau)^2 [ -d\tau^2 + \{ (1-2\mathcal{R})\delta_{ij}
+ h_{ij}\} dx^i dx^j],
\end{align}
where $\mathcal{R}$ represents the curvature perturbation
on the uniform field hypersurfaces and $h_{ij}$ is the tensor perturbation
that satisfies  $h_i^i = 0 = {h^i}_{j,i}$.

If we perform the Fourier transform of $\mathcal{R}$ and $h_{ij}$,
\begin{align}
\mathcal{R}(\tau, {\bf x}) =&  \frac{1}{z_s}
\int \frac{d^3k}{(2\pi)^{3/2}}
{v}_s(\tau,k) e^{i{\bf k}\cdot {\bf x}}, \\
h_{ij} (\tau, {\bf x}) =& \frac{2}{z_t}\sum_{\lambda}
\int \frac{d^3k}{(2\pi)^{3/2}}
v^{\lambda}_t(\tau,k) \epsilon_{\lambda,ij} e^{i{\bf k}\cdot{\bf x}},
\label{ftt}
\end{align}
where $\epsilon_{ij}$ is a polarization tensor, Sasaki-Mukhanov equations for $v_s$ and $v_t$ are
derived from linearizing Eqs. (\ref{einstein})--(\ref{kg})
\begin{align}
& v_s'' + \left(c_s^2 k^2 - \frac{z''_s}{z_s}\right)v_s = 0,
\label{sms} \\
& v_t'' + \left(c_t^2 k^2 -\frac{z_t''}{z_t} \right) v_t = 0,
\label{smt}
\end{align}
where \cite{Guo:2010jr}\cite{Hwang:2005hb}
\begin{align}
z_s &\equiv \sqrt{\frac{a^2
(\dot{\phi}^2 +  6\dot{\xi} H^3 \Delta )}
{ H^2 (1-\frac{1}{2}\Delta)^2}} ,
\quad \Delta = \frac{4\kappa^2 \dot{\xi} H}{1-4\kappa^2 \dot{\xi}H}, \\
z_t &\equiv \sqrt{\frac{a^2}{\kappa^2}(1 -4\kappa^2\dot{\xi}H)},
\end{align}
and
\begin{align}
c_s^2 \equiv & 1+\frac{2(\dot{H}-\kappa^2\dot{\xi}H
(H^2 +4\dot{H}) +\kappa^2 \ddot{\xi}H^2)\Delta^2}{\kappa^2\dot{\phi}^2
+6\kappa^2\dot{\xi}H^3 \Delta},
\\
c_t^2 \equiv & 1 - \frac{4\kappa^2(\ddot{\xi}-\dot{\xi}H)}
{1-4\kappa^2\dot{\xi}H}.
\end{align}
Here, a prime represents a derivative with respect
to the conformal time $\tau = \int a^{-1}dt$.

 $z_A$ and  $c_A^2$, where
$A=\{s,t\}$, can be written in terms of the slow-roll parameters
\cite{Guo:2010jr}\cite{Hwang:2005hb}
using the definitions of the slow-roll parameters (\ref{srparam}):
\begin{align}
z_s =& \sqrt{\frac{a^2}{\kappa^2}
\frac{2\epsilon -\delta_1(1+2\epsilon-\delta_2)
+\frac{3}{2}\delta_1 \Delta}{(1-\frac{1}{2}\Delta)^2}},
\quad \Delta = \frac{\delta_1}{1-\delta_1}
\label{zssr},
\\
z_t =& \sqrt{\frac{a^2}{\kappa^2}(1-\delta_1)}
\label{ztsr},
\\
c_s^2 =& 1 - \frac{ (4 \epsilon
+ \delta_1 (1- 4\epsilon -\delta_2) )\Delta^2}{4\epsilon - 2\delta_1
-2\delta_1 (2\epsilon -\delta_2) +3 \delta_1 \Delta},
\\
c_t^2 =& 1 + \frac{\delta_1 (1 - \delta_2)}{1-\delta_1},
\end{align}
where we have used the following relation from Eqs. (\ref{beq2})--(\ref{beq3}):
\begin{align}
\frac{\kappa^2 \dot{\phi}^2}{H^2} = 2\epsilon - \delta_1
(1+2\epsilon -\delta_2).
\end{align}
If one keeps the leading order of the slow-roll parameters in $z_A''/z_A$
using (\ref{zssr})--(\ref{ztsr}),
Eqs. (\ref{sms})--(\ref{smt}) become
\begin{align}
& v_A'' + \left( c_A^2 k^2 - \frac{\nu_A^2 -1/4}{\tau^2} \right)
v_A =0,
\label{smlead}
\end{align}
where  the parameters are given by up to leading order in slow-roll parameters
\begin{align}
\nu_s &\simeq \frac{3}{2} + \epsilon +
\frac{2\epsilon (2\epsilon +\eta) - \delta_1 (\delta_2 -\epsilon)}
{4\epsilon-2\delta_1}
\\
\nu_t &\simeq \frac{3}{2} +\epsilon.
\end{align}
In deriving (\ref{smlead}), we use the following relation:
\begin{align}
\tau &= -\frac{1}{aH} \frac{1}{1-\epsilon}.
\label{conformalsr}
\end{align}
One  can obtain the exact solutions for (\ref{smlead}) assuming that the
slow-roll parameters are constants,
\begin{align}
v_{A} &= \frac{\sqrt{\pi|\tau|}}{2}[ c^A_1(k)
H_{\nu_{A}}^{(1)}(c_{A} k|\tau|) + c^A_2 (k) H_{\nu_{A}}^{(2)}
(c_{A}k|\tau|)],
\label{exactsol}
\end{align}
where  $H_{\nu}^{(i)}\,\, (i=1,2)$ are the first and
second kind Hankel functions.
 $c^A_i\,\, (i=1,2)$ are the
coefficients that are determined from the initial conditions
and satisfy the normalization conditions
\begin{align}
|c^A_2|^2 - |c^A_1|^2 = 1.
\end{align}

If we adopt the Bunch-Davies vacuum for the initial fluctuation modes
at $ c_A k|\tau| \gg 1$ by taking the positive mode frequency,
the initial modes are given by
\begin{align}
v_A = \frac{1}{\sqrt{2 c_A k}} e^{i c_A k |\tau|}.
\label{ic}
\end{align}
These modes correspond to the choice of the coefficients
\begin{align}
c_1^{A} =  e^{i(\nu_A+\frac{1}{2})\frac{\pi}{2}},
\quad c_2^{A} = 0,
\end{align}
where we have used the asymptotic form of the
 Hankel functions in the limit $x \equiv c_A k|\tau| \gg 1$,
\begin{align}
H_{\nu_A}^{(1,2)}(x) \sim \sqrt{\frac{2}{\pi x}} e^{\pm
i(x-(\nu_A +\frac{1}{2} )\frac{\pi}{2}) }.
\end{align}
Then the exact solution (\ref{exactsol}) becomes
\begin{align}
v_A = \frac{\sqrt{\pi |\tau|}}{2} e^{i(\nu_A+\frac{1}{2})\frac{\pi}{2}}
H_{\nu_A}^{(1)} (c_A k|\tau|).
\label{exactsol2}
\end{align}

The power spectra  of the scalar and tensor modes are calculated with
(\ref{exactsol2}) on the large scales.
Since the first kind Hankel function is approximated
in the large scale limit ($c_A k |\tau| \ll 1$) as
\begin{align}
H_{\nu_A}^{(1)} \sim \frac{2}{1-e^{2i\nu_A \pi}}
\biggl\{ \frac{1}{\Gamma(1+\nu_A)} \left(\frac{x}{2}\right)^{\nu_A}
-\frac{e^{i\nu_A \pi}}{\Gamma(1-\nu_A)}
\left(\frac{x}{2} \right)^{-\nu_A} \biggr\},
\end{align}
where the second term is dominant, one can obtain
the power spectra for the scalar and tensor modes
on the large scales
\begin{align}
\mathcal{P}_s &=
\frac{k^3}{2\pi^2} \left|\frac{v_s}{z_s}\right|^2
\nonumber \\
&\simeq \frac{\csc^2 \nu_s \pi}{\pi \mathcal{D}_s^2
\Gamma^2(1-\nu_s)} \frac{1}{c_s^3 |\tau|^2 a^2 }
\biggl(\frac{c_s k|\tau|}{2}\biggr)^{3-2\nu_s},
\label{psfs}
\\
\mathcal{P}_t &= 2\frac{k^3}{2\pi^2} \left|\frac{2v_t}{z_{t}}\right|^2
\nonumber \\
&\simeq 8\frac{\csc^2 \nu_t \pi}{\pi \mathcal{D}_t^2
\Gamma^2(1-\nu_t)} \frac{1}{c_t^3 |\tau|^2 a^2 }
\biggl(\frac{c_t k|\tau|}{2}\biggr)^{3-2\nu_t},
\label{psft}
\end{align}
where the factor 2 of the tensor power spectrum comes from
the two polarization states and we define
\begin{align}
z_A &\equiv \mathcal{D}_A a^2,
\nonumber \\
\mathcal{D}_s^2 &=\frac{2\epsilon
-\delta_1(1+2\epsilon-\delta_2)
+\frac{3}{2}\delta_1 \Delta}{\kappa^2 (1-\frac{1}{2}\Delta)^2},
\nonumber \\
\mathcal{D}_t^2 &= \frac{1-\delta_1}{\kappa^2}.
\nonumber
\end{align}

The spectral indices of the scalar and tensor modes and the  tensor-to-scalar ratio are
given by
\begin{align}\label{specind}
n_s -1 &\equiv  \frac{d\ln
\mathcal{P}_{s}}{d\ln k}
\nonumber \\
&= 3-2\nu_{s} \approx -2\epsilon
- \frac{2\epsilon (2\epsilon +\eta) -\delta_1 (\delta_2 -\epsilon)}
{2\epsilon - \delta_1},
\\
n_t &\equiv \frac{d\ln \mathcal{P}_t}{d\ln k}
=3-2\nu_t \approx -2\epsilon,
\\
r &\equiv \frac{\mathcal{P}_t}{\mathcal{P}_s}
\approx  8 (2\epsilon -\delta_1). \label{ttsr}
\end{align}
We can also calculate the running  spectral indices  of the  scalar and tensor modes
\begin{align}
\frac{d n_s}{d\ln k} \approx& -2\epsilon (2\epsilon+ \eta)
 +\frac{(2\epsilon (2\epsilon +\eta) -\delta_1 (\delta_2 - \epsilon))^2}
{(2\epsilon-\delta_1)^2}
\nonumber \\
& - \frac{2\epsilon(8\epsilon^2 +7\epsilon \eta + \zeta)
+\delta_1 (\epsilon^2 +\epsilon\eta + \epsilon \delta_2
-\delta_3)}{2\epsilon -\delta_1}
\label{dns} \\
\frac{d n_t}{d\ln k} \approx& -2 (2\epsilon^2 + \epsilon \eta).
\label{dnt}
\end{align}
where we have used from (\ref{srparam})
\begin{align}
\frac{d\epsilon}{d\ln k} &= 2\epsilon^2 + \epsilon \eta,
\\
\frac{d\eta}{d\ln k} &= \epsilon \eta - \eta^2 + \zeta,
\\
\frac{d\delta_1}{d\ln k} &= \delta_1 (\delta_2 - \epsilon),
\\
\frac{d\delta_2}{d\ln k} &= \epsilon \delta_2 - \delta_2^2 +\delta_3.
\end{align}

\section{Models}\label{sect4}

In this section, we calculate the $n_s, \,\, n_t,\,\,r$, and $\frac{dn_s}{d\ln k}$
for the specific models using Eqs. (\ref{specind})--(\ref{dns}) and then constrain  our model predictions  with the recent CMB observational data from
Planck and BICEP2.

\subsection{Exponential potential with an exponential Gauss-Bonnet coupling}
\label{expot_expcoup}

Let us start with the exponential potential and  exponential coupling to the GB term
\bea
\label{expotential}
V(\phi)= V_0 e^{-\lambda \phi}, \quad \xi(\phi)=\xi_0 e^{-\lambda \phi},
\ena
where $V_0,\xi_0$ and $\lambda$ are constants. One can calculate the slow-roll parameters, (\ref{srpote})--(\ref{srpotd}), for the model
given by (\ref{expotential})
\bea \label{srp1}
\e &=& \frac{1}{2}\lambda^2 e^{-2 \lambda \phi } \left(\alpha +e^{2 \lambda  \phi }\right), \\
\eta &=& -\lambda ^2 \left(3 \alpha  e^{-2 \lambda  \phi }+1\right), \\
\d_1 &=& - \alpha  \lambda ^2 e^{-4 \lambda  \phi } \left(\alpha +e^{2 \lambda  \phi }\right), \\
\d_2 &=& -\frac{1}{2} \lambda ^2 \left(7 \alpha  e^{-2 \lambda  \phi }+3\right).
\label{srp4}
\ena
Inflation ends at $\epsilon(\phi_e)=1$, although inflation does not stop naturally for this scenario, which gives the value of the field at the end of the inflation
\bea \label{phiend}
\phi_e = -\frac{1}{2\l} \ln \left(\frac{2-\lambda ^2}{\alpha  \lambda ^2}\right),
\ena
where $\lambda\neq0$. In this section we consider that the value of the field at the end of inflation is much smaller than that of the beginning, which means $\phi_e\ll\phi$.
Therefore, the number of $e$-folds before the end of inflation is
\bea \label{ne2}
N \simeq \int_{\phi_e}^{\phi}\frac{\kappa^2}{Q} d\phi
= -\frac{1}{2 \l^2}\ln \left(\alpha +e^{2 \lambda  \phi }\right).
\ena
From (\ref{ne2}), we obtain
\bea \label{phi2}
\phi =\frac{1}{2 \lambda } \ln \left(e^{-2 \lambda ^2 N}-\alpha \right).
\ena
After substituting the last result (\ref{phi2}) into (\ref{specind}) and (\ref{ttsr}), the spectral index of the scalar modes $n_s$ and tensor-to-scalar ratio $r$ can be written as
\bea \label{pl1}
n_s-1 &=& \lambda ^2 \left(\frac{3 \alpha }{e^{-2 \lambda ^2 N}-\alpha }-1\right),
\quad
r = \frac{8 \lambda ^2 e^{-4 \lambda ^2 N}}{\left(e^{-2 \lambda ^2 N}-\alpha \right)^2}.
\ena
One, then, can write the relation between $r$ and $n_{S}$ as follows:
\bea \label{pol3}
r=-\frac{8}{4 \alpha ^2 e^{4 \lambda ^2 N}-5 \alpha
e^{2 \lambda ^2 N}+1}(n_s-1).
\ena
Before we compare our theoretical predictions with the observational data by Planck,
one last thing that we need to check is the valid model parameter ranges
for inflation to happen.

From (\ref{phiend})--(\ref{phi2}) , we find that
\begin{align} \label{con1}
& \frac{2-\l^2}{\a \l^2}>0 ,
\quad 0<\alpha + e^{2 \lambda \phi} <1,
\quad \mbox{and} \quad e^{-2\lambda^2 N} > \alpha.
\end{align}
Since  $\l^2$ is always positive ($\lambda^2 > 0$)
and $\alpha$ can be negative or positive,
we can reach to the following results: if $\alpha> 0$,
$0 < \alpha< e^{-2\lambda^2 N}$, then $-\sqrt{2} <\lambda <\sqrt{2}$.
Or if $\alpha < 0$, 
then  $\lambda < -\sqrt{2} \,\, \text{or} \,\, \lambda > \sqrt{2}$.
With these parameter ranges, we can freely choose the model parameters $\alpha$ and $\lambda$ that are
valid for  inflation to occur. Unfortunately, these parameter
ranges of $\alpha$ and $\lambda$ are not favored by observational data.


\subsection{Power-law potential and  power-law Gauss-Bonnet coupling}
\label{powpot_powcoup}

We consider an inflationary model
with the power-law  potential and  power-law coupling
to the Gauss-Bonnet term  characterized as follows:
\bea
\label{powpot}
V(\phi) = V_0 \phi^n, \quad \xi(\phi) = \xi_0 \phi^{n}.
\ena
This class of potential has been widely studied as a simplest inflationary model and includes the simplest chaotic models, in which inflation starts from the large values of an inflaton field,
$\phi>M_p$.

For the model with the choice of (\ref{powpot}), the slow-roll parameters
can be calculated using (\ref{srpote})--(\ref{srpotd}) as
\begin{align} \label{genp1}
\e \simeq &\frac{n^2}{2\kappa^2}(1 + \alpha \phi^{2n})\phi^{-2}, \\
\eta \simeq& -\frac{n}{\kappa^2} \left[n-2
+ (3 n-2)\alpha \phi^{2 n}\right]\phi^{-2}, \\
\zeta \simeq& \frac{n^2}{2\kappa^4}\biggl[
16-14n+3n^2 + 4(n-1)(7n-8) \alpha \phi^{2n}
\nonumber \\
& + (3n-2)(11n-8) \alpha^2 \phi^{4n} \biggr] \phi^{-4},
\\
\d_1 \simeq& -\frac{n^2}{\kappa^2} \alpha \phi^{2n} (1
+ \alpha \phi^{2n})\phi^{-2}, \\
\d_2 \simeq& -\frac{n}{2\kappa^2} \left(3 n-4 + (7 n-4) \alpha  \phi ^{2 n}\right)\phi^{-2},\label{genp2}
\\
\delta_3 \simeq& \frac{n^2}{\kappa^4} \biggl[
8-10n +3n^2 +4 (n-1)(5n-4)\alpha \phi^{2n}
\nonumber \\
& +(3n-2)(7n-4)\alpha^2 \phi^{4n} \biggr]\phi^{-4}.
\end{align}
The number of $e$-folds before the end of inflation  for the
choices of (\ref{powpot}) is given in
(\ref{efold_pow}) by
\bea \label{gen2}
N 
\simeq \frac{\kappa^2 \phi^2}{2n}
\, _2F_1\left(1;\frac{1}{n};1+\frac{1}{n};-\alpha \phi ^{2 n}\right).
\nonumber
\ena
It turns out that ${}_{2}F_{1}\left(1;\frac{1}{n};1+\frac{1}{n};0\right)=1$ for $\alpha=0$;
then we can reproduce the standard chaotic inflation results,
$\kappa^2 \phi^2=2nN$. Here, we assume the term of $-\a \phi^{2n}$ to be much smaller than 1,
so that we could expand the hypergeometric function up to
the leading order in $\alpha$,
\begin{align}
{}_{2}F_{1} \left(1;\frac{1}{n}; 1+\frac{1}{n};-\alpha \phi^{2n} \right) \approx
1-\frac{\alpha \phi^{2n}}{n+1} + \mathcal{O}(\alpha^2).
\end{align}
 Then the number of $e$-folds  becomes
\bea \label{gen3}
N \simeq \frac{\kappa^2 \phi ^2}{2 n}
\left(1-\frac{\alpha \phi ^{2 n}}{n+1}\right)
+\mathcal{O}(\alpha^2).
\ena
As we described in Sec. \ref{sect2}, $\alpha \lesssim 10^{-6}\,\, M_p^{-4}$
for $n=2$
and $\alpha \lesssim 10^{-12}\,\, M_p^{-8}$ for $n=4$ to
have enough $e$-folding, $N \gtrsim 60$. This implies that $\alpha$ can be treated
as a small parameter.

We also expand $\phi$ to the leading order in $\tilde{\alpha}$,
which is a dimensionless parameter, $\tilde{\alpha} = \alpha M_p^{2n}$,
\begin{align}
\phi=\phi^{(0)}+\tilde{\a}\phi^{(1)}+\mathcal{O}(\tilde{\alpha^2}).
\label{expnphi}
\end{align}
Substituting (\ref{expnphi})  into (\ref{gen3}), we obtain
\bea \label{gen4}
\phi \simeq \sqrt{\frac{2 n N}{\kappa^2}} \left[1+\frac{\a(2 n N)^n}{2(n+1)\kappa^{2n}}\right].
\ena
With  (\ref{gen4}), one can rewrite (\ref{genp1})--(\ref{genp2}) as follows:
\begin{align}\label{genp11}
\epsilon \simeq& \frac{n}{4 N} +
\frac{n^2(2nN)^n \alpha}{4(1+n)N\kappa^{2n}},
\\
\eta \simeq& \frac{2-n}{2N}
-\frac{ 3n^2 (2nN)^n \alpha}{2(1+n)N\kappa^{2n}},
\\
\zeta \simeq& \frac{(n-2)(3n-8)}{8N^2}
+\frac{n^2 (14n-19) (2nN)^n \alpha}{4(1+n)N^2\kappa^{2n}},
\\
\delta_1 \simeq& - \frac{n(2nN)^n \alpha}{2N \kappa^{2n}},
\\
\delta_2 \simeq& \frac{4-3n}{4N}
- \frac{7n^2 (2nN)^n \alpha}{4(1+n)N\kappa^{2n}},
\\
\delta_3 \simeq& \frac{(n-2)(3n-4)}{4N^2}
+ \frac{n^2 (10n-11) (2nN)^n \alpha}{2(1+n)N^2\kappa^{2n}} \label{genp22}.
\end{align}

Substituting (\ref{genp11})--(\ref{genp22})
 into  (\ref{specind})--(\ref{dnt}), we obtain $n_s,\,\,r,
\,\,n_t, \,\, \frac{dn_s}{d\ln k}$, and
$\frac{dn_t}{d\ln k}$, respectively, as follows:
\begin{align}
n_s -1 \simeq& -\frac{n+2}{2 N} +
\frac{ n(3n+2)  (2nN)^n \alpha}{2(1+n)N\kappa^{2n}},
\\
n_t \simeq& -\frac{n}{2N} -\frac{n^2(2nN)^n \alpha}{2(1+n)N\kappa^{2n}},
\\
r \simeq& \frac{4n}{N}+ \frac{4n(2n+1) (2nN)^n
\alpha}{(1+n)N\kappa^{2n}},
\\
\frac{dn_s}{d\ln k}\simeq& -\frac{n+2}{2N^2}
-\frac{n(n-1)(3n+2) (2nN)^n\alpha}
{2(1+n) N^2 \kappa^{2n}},
\label{dns_pow}
\\
\frac{dn_t}{d\ln k} \simeq& -\frac{n}{2N^2}
+\frac{n^2(n-1) (2nN)^n\alpha}{2(1+n)N^2\kappa^{2n}}.
\end{align}

\begin{figure}[tbp]
  \centering
    \includegraphics[width=0.5\textwidth]{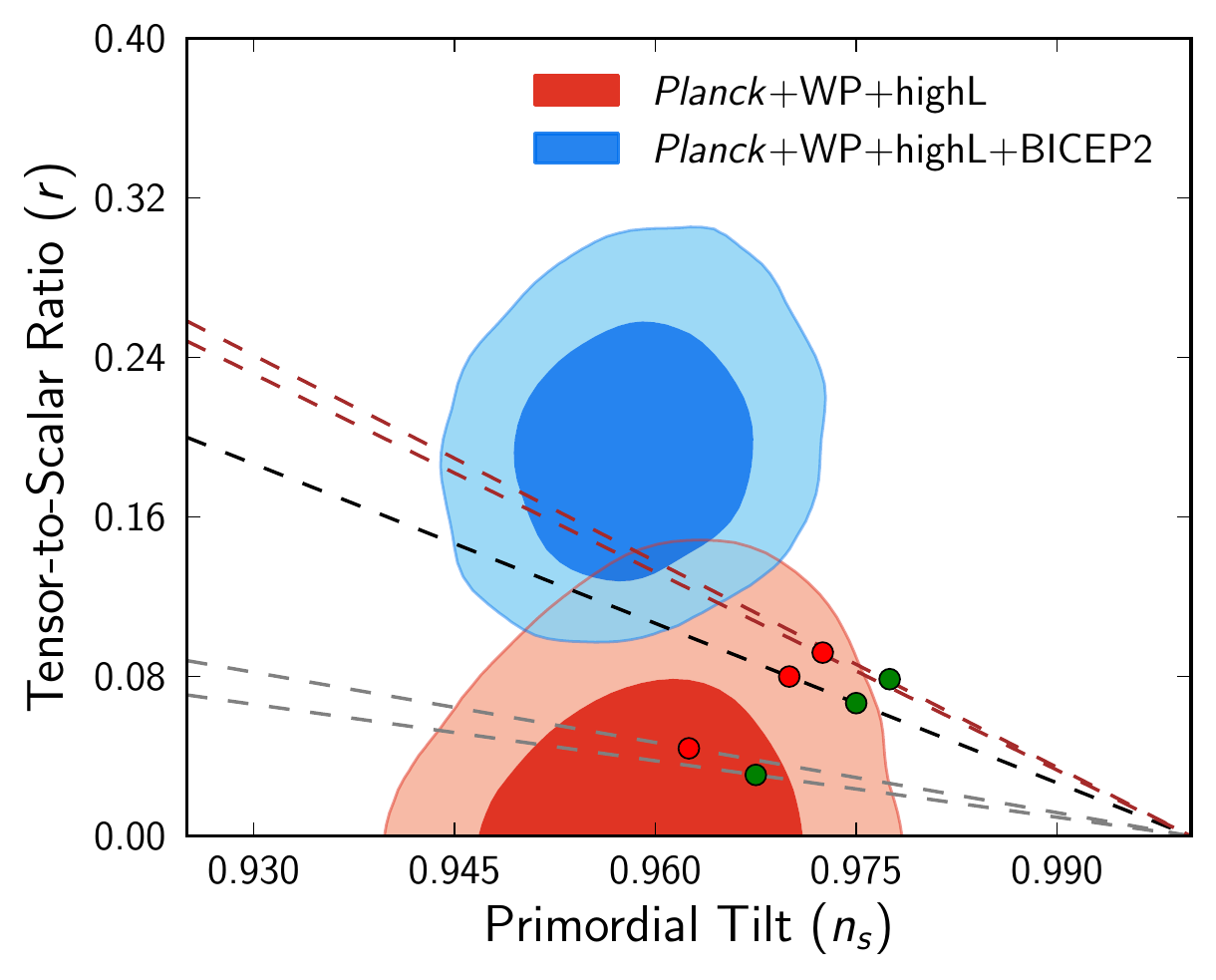}
      \caption{\small Marginalized joint $68\%$ and $95\%$ CL regions for ($n_s$, $r$), using observational data sets with and without a running spectral index, compared to the theoretical prediction of the model (\ref{powpot}) with $n=1$. The black dashed line is for the case where model parameter $\alpha=0$ while gray and brown are for the case where $\alpha=-0.003$ and $\alpha=0.001$, respectively. The pairs of red and green dots represent the number of $e$-folds, $N=50$ and $N=60$, respectively.}
\label{fig_nsrn1}
\end{figure}

\begin{figure}[tbp]
  \centering
      \includegraphics[width=0.5\textwidth]{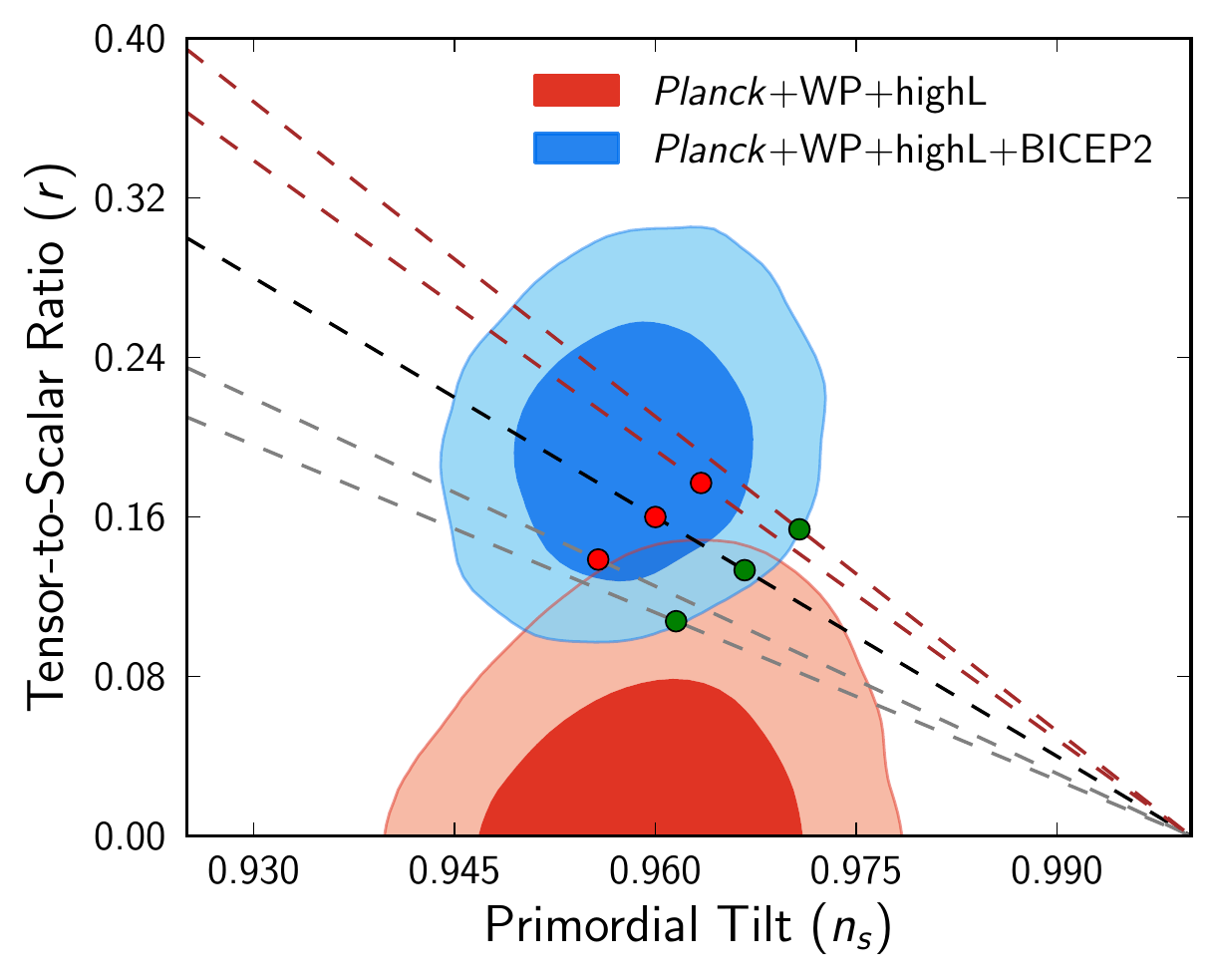}
      \caption{{\small Marginalized joint $68\%$ and $95\%$ CL regions for ($n_s$, $r$), using observational data sets with and without a running spectral index, compared to the theoretical prediction of the model (\ref{powpot}) with $n=2$. The black dashed line is for the case where model parameter $\alpha=0$ while gray and brown are for the case where $\alpha=-2\times 10^{-6}$ and $\alpha=1.5\times 10^{-6}$, respectively. The pairs of red and green dots represent the number of $e$-folds, $N=50$ and $N=60$, respectively.}}
\label{fig_nsrn2}
\end{figure}

\begin{figure}[tbp]
\centering
    \includegraphics[width=0.5\textwidth]{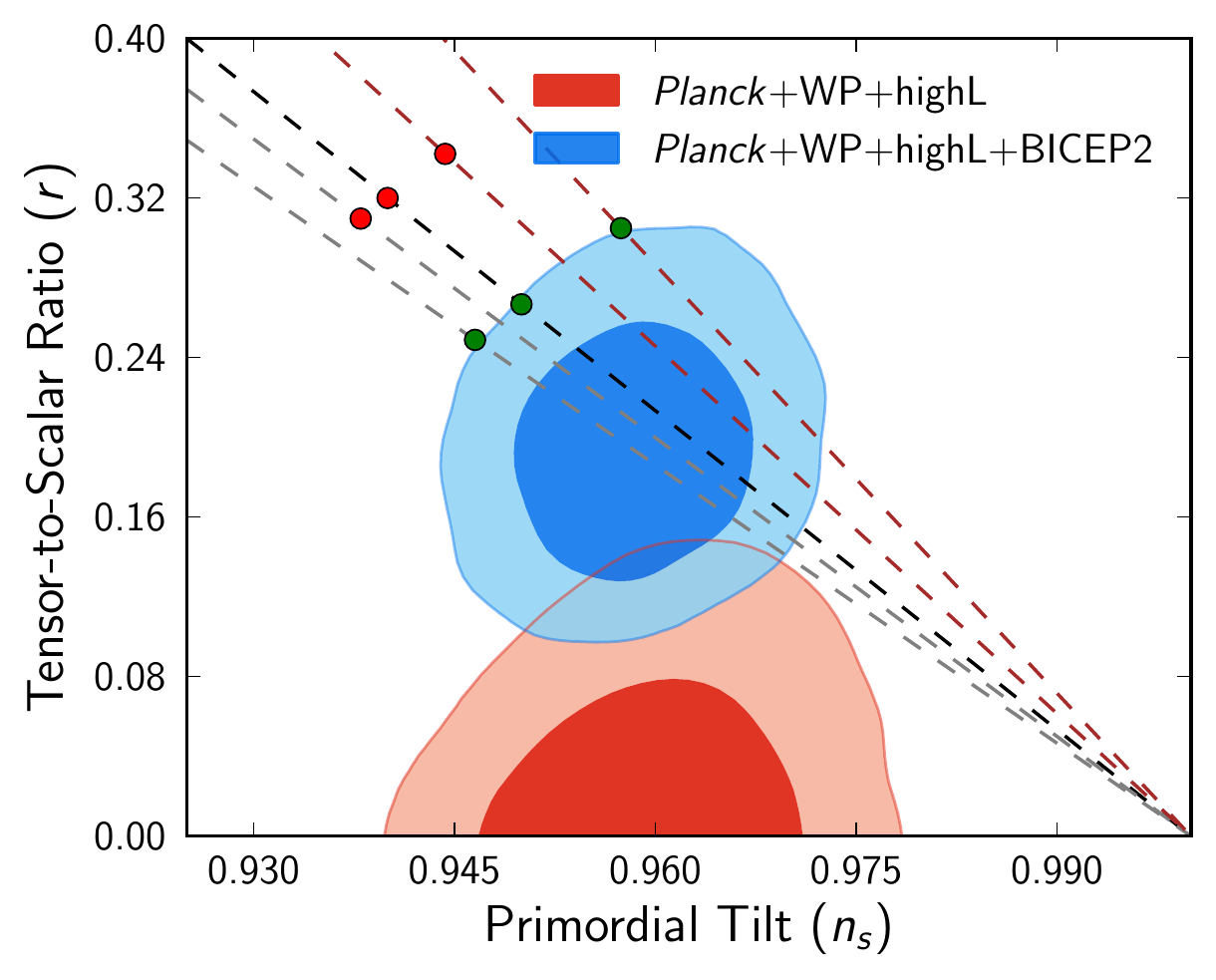}
      \caption{{\small Marginalized joint $68\%$ and $95\%$ CL regions for ($n_s$, $r$), using observational data sets with and without a running spectral index, compared to the theoretical prediction of the model (\ref{powpot}) with $n=4$. The black dashed line is for the case where model parameter $\alpha=0$ while gray and brown are for the case where $\alpha=-0.7\times10^{-12}$ and $\alpha=1.5\times 10^{-12}$, respectively. The pairs of red and green dots represent the number of $e$-folds, $N=50$ and $N=60$, respectively.}}
\label{fig_nsrn4}
\end{figure}

Figures \ref{fig_nsrn1}--\ref{fig_nsrn4}
show the $n_s$-$r$ contour plot of the models that are  given by
(\ref{powpot}) with $n=1$, $n=2$, and $n=4$
for the  different values of $N$ and $\a$
in  comparison with the observational data.
The red contour comes from the  Planck data
and the BICEP2 data set are included
in the blue contour. The Planck and WMAP data constrain on $r$ as
$r < 0.12$, but BICEP2 claims that $r \simeq 0.2$. There seems to
be some discrepancy between Planck and BICEP2. One way out of
this discrepancy might be to take into account the running spectral
index of the scalar modes \cite{Ade:2014xna}.

Black, brown, and gray dashed lines represent the theoretical predictions
for $\alpha = 0$ (black), $\alpha > 0$ (brown), and $\alpha < 0$ (gray), respectively,
and the pairs of red and blue dots represent $N=50$ and $N=60$, respectively,
in Figs. \ref{fig_nsrn1}--\ref{fig_nsrn4}.

Without the Gauss-Bonnet term $(\alpha =0$), Planck data  say that
the $\phi^4$ model lies well outside of the joint 99.7\% CL (confidence level)
region in the $n_s-r$ plane (Fig. \ref{fig_nsrn4}) and the $\phi^2$ model
lie outside of the 95\% CL region for $N\lesssim 50$ (Fig. \ref{fig_nsrn2}).
On the contrary, the inflationary models with $n=1$ lies within the $95\%$ CL regions
(Fig. \ref{fig_nsrn1}).
If we consider the combination of BICEP2 and Planck,
even $N=60$ for $\phi^4$ reside within
the $95\%$ CL regions, but $n=1$ model might be ruled out.

Both  $n_s$ and $r$ are suppressed  if $\alpha \neq 0$
and  has negative values, but, for $\alpha > 0$, those are enhanced.  These
results are completely opposite compared to Ref. \cite{Guo:2010jr}, in which
$r$ is  enhanced for negative $\alpha$ and reduced for positive $\alpha$ for $V=V_0 \phi^n$ with $\xi=\xi_0 \phi^{-n}$. Because $r$ becomes suppressed as $n$ decreases,
Planck  data alone favor the $n=1$ model, but the BICEP2 + Planck
favors $n=2$.  Even for $\alpha \neq 0$, BICEP2 with Planck seems to
rule out $n=1$ at $95\%$ CL (Fig. \ref{fig_nsrn1}).
For $n=2$ with $\alpha \neq 0$ (Fig. \ref{fig_nsrn2}),
negative $\alpha$ with $N=60$ lies within the contour of $95\%$ CL according to Planck data,
but positive $\alpha$ is located outside of the contour.
We find that Planck alone favors $\alpha < 0$ model with $N> 50$.
However,  Planck combining with BICEP2 allows both the positive and negative
$\alpha$ models with $N=50$ and $N=60$  at $95\%$ CL and favors
$N \lesssim 60$. Although the  $n=4$ model seems to be ruled out
by Planck at $99.7\%$ CL \cite{Ade:2013uln}, it can be survived  according to Planck combining
by BICEP2 (Fig. \ref{fig_nsrn4}) for $N> 50$ at $95\%$ CL.

To note, data do not constrain on $n_t$; therefore, in our analysis, $n_t$ is varied independent of the tensor-to-scalar ratio \cite{Guo:2010jr}.

In Table \ref{tab4}, we list the range of the model parameter
$\a$ in which the predicted value of $n_s$ and $r$ is consistent
with Planck + BICEP2.
\begin{table}[tbp]
\caption{Observationally favored range of model parameters $\a$ for different values of $n$ and $N$ from the observational data set.}
\centering 
\begin{tabular}{c c c } 
\hline\hline 
Model & Parameter range & Parameter range \\
$n$ & for N=50 & for N=60 \\ [0.5ex]
\hline 
n=1   & $-6.6\times 10^{-3}\leq\a\leq2\times 10^{-3}$ & $-5.5\times10^{-3} \leq\a\leq 4\times10^{-4}$ \\
n=2   & $-5.2\times 10^{-6}\leq\a\leq6\times 10^{-6}$ & $-3.2\times10^{-6} \leq\a\leq 1.5\times10^{-6}$ \\
n=4   & ${\small \mbox{lies outside of } 2\sigma \mbox{ boundary}} $ & $ -0.7\times10^{-12} \leq\a\leq 1.5\times10^{-12}$\\
\hline 
\end{tabular}
\label{tab4} 
\end{table}

\begin{figure}[tbp]
  \centering
    \includegraphics[width=0.8\textwidth]{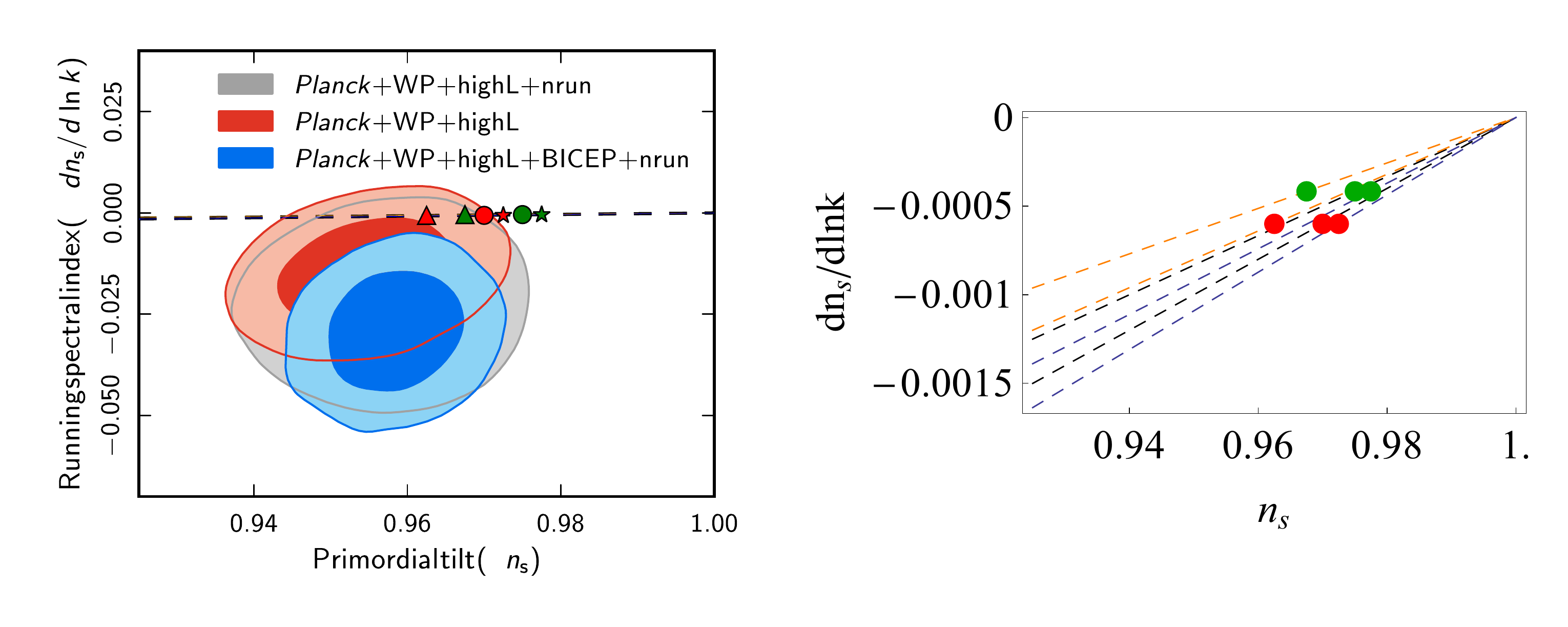}
      \caption{\small $n_s$ vs $dn_{s}/d\ln k$ plot, with observational data sets (left) and without observational data sets (right), compared to the theoretical prediction of the model (\ref{powpot}) with $n=1$. The black dashed line is for the case where model parameter $\alpha=0$ while orange and blue are for the case where $\alpha=-0.003$ and $\alpha=0.001$, respectively. The pairs of red and green markers represent the number of $e$-folds, $N=50$ and $N=60$, respectively.}
      \label{fig_dnsn1}
\end{figure}

\begin{figure}[tbp]
  \centering
      \includegraphics[width=0.8\textwidth]{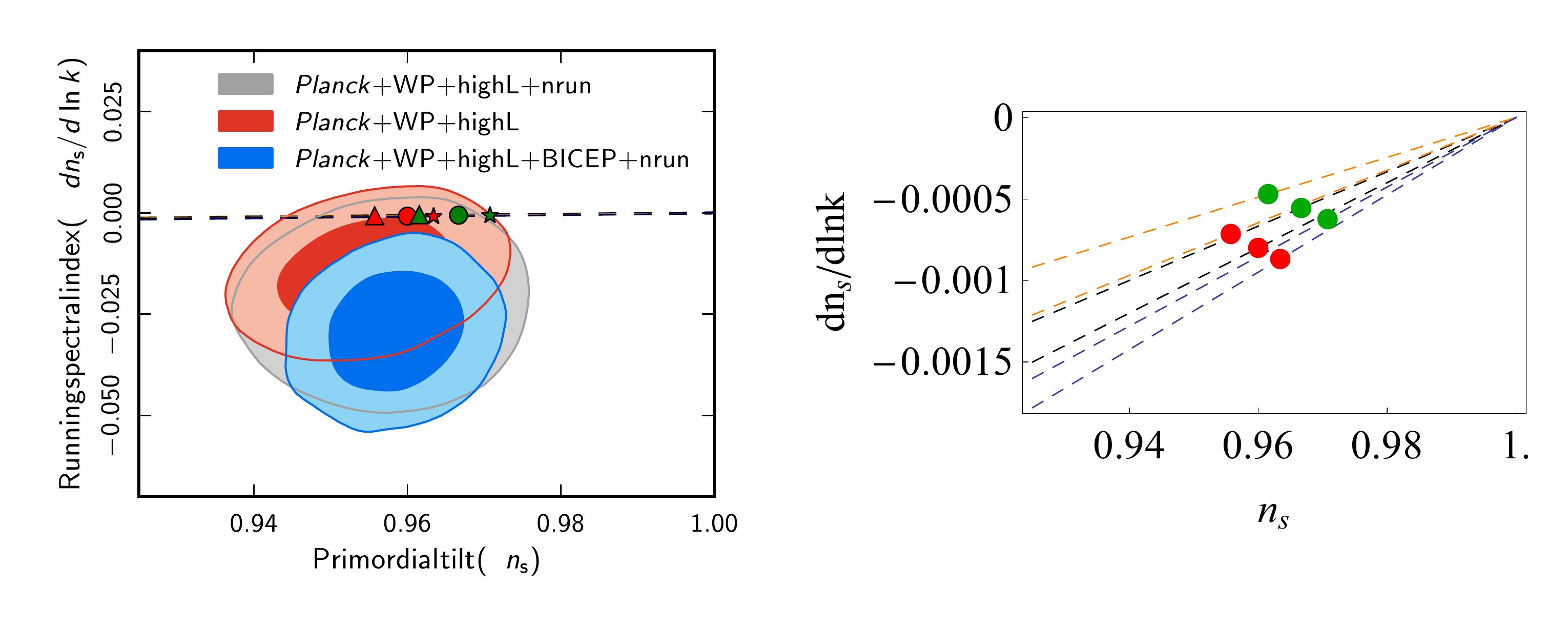}
      \caption{{\small $n_s$ vs $dn_{s}/d\ln k$ plot, with observational data sets (left) and without observational data sets (right), compared to the theoretical prediction of the model (\ref{powpot}) with $n=2$. The black dashed line is for the case where model parameter $\alpha=0$ while orange and blue are for the case where $\alpha=-2\times 10^{-6}$ and $\alpha=1.5\times 10^{-6}$, respectively. The pairs of red and green markers represent the number of $e$-folds, $N=50$ and $N=60$, respectively.}}
      \label{fig_dnsn2}
\end{figure}

\begin{figure}[tbp]
\centering
    \includegraphics[width=0.8\textwidth]{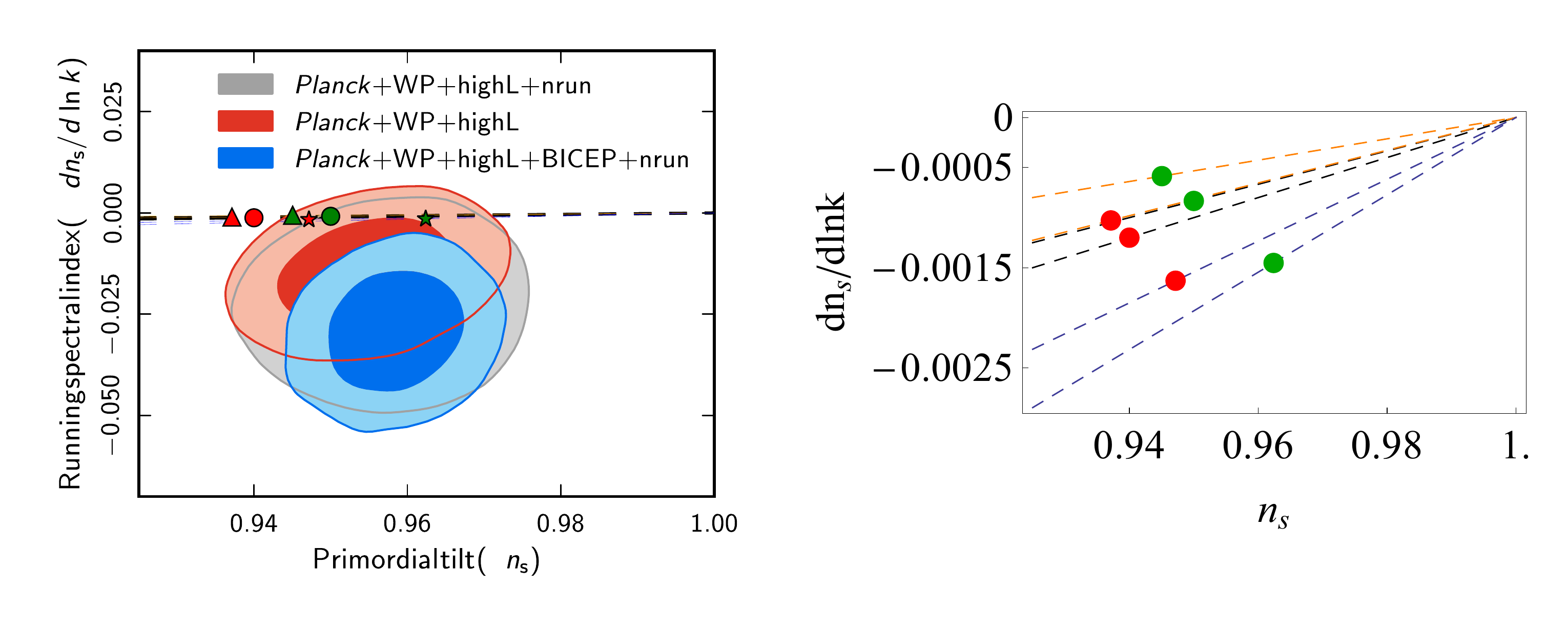}
      \caption{{\small $n_s$ vs $dn_{s}/d\ln k$ plot, with observational data sets (left) and without observational data sets (right), compared to the theoretical prediction of the model (\ref{powpot}) with $n=4$. The black dashed line is for the case where model parameter $\alpha=0$ while orange and blue are for the case where $\alpha=-10^{-12}$ and $\alpha=2.5\times 10^{-12}$, respectively. The pairs of red and green markers represent the number of $e$-folds, $N=50$ and $N=60$, respectively.}}
      \label{fig_dnsn4}
\end{figure}

We plot the $dn_s/d\ln k$ with $n_s$ in Figs. \ref{fig_dnsn1}--\ref{fig_dnsn4}.
The theoretical prediction from (\ref{dns_pow}) is
$|\frac{dn_s}{d\ln k}| \sim  10^{-4}$, but Planck combined with the BICEP2 data provides
$|\frac{dn_s}{d\ln k}| \sim  10^{-2}$, which is
larger than the theoretical predictions.
 Therefore, the predictions of
our model lie outside of $95\%$ regions. Meanwhile, the model
predictions  are consistent with Planck alone with or without
the running spectral index at $95\%$ CL.  We may understand this as follows:
As we explained in Sec. \ref{intro}, we consider the running spectral index of the scalar modes
as a simple resolution  to  reconcile BICEP2 with  Planck, which was considered in Ref. \cite{Ade:2014xna}.
Either our theoretical model is not consistent with the observations under the
assumption that  the BICEP2 data is correct or considerations of the running spectral index
may not be the right resolution to reconcile both data.

Therefore, we need to find another way to resolve the inconsistency with BICEP2 data
since our predictions are not consistent with the $dn_s/d\ln k$-$n_s$ data.
We will leave it for a future work because these considerations
are beyond the scope of this work.


\section{Conclusions and discussions} \label{summary}

We have investigated the slow-roll inflation
with the Gauss-Bonnet term that is coupled to the  inflaton field nonminimally.
We have considered the potential and  coupling function
as $V(\phi) = V_0 e^{\lambda \phi},\,\,
\xi(\phi) =\xi_0 e^{\lambda \phi}$ (Sec. \ref{expot_expcoup}) and
$V(\phi) = V_0 \phi^n, \,\, \xi(\phi) = \xi_0 \phi^n$
(Sec. \ref{powpot_powcoup}), respectively, to relax the condition
$V(\phi) \xi(\phi) = {\rm const.}$, which was widely studied
in Refs. \cite{Guo:2010jr}--\cite{Jiang:2013gza}.
$N \simeq 60$ condition requires that $\alpha \simeq 10^{-6}$ for
$V \sim \phi^2$ and $\alpha \simeq 10^{-12}$ for $ V \sim \phi^4$
(see Fig. \ref{fig_efolding}) where
$\alpha = \frac{4}{3} \kappa^4 V_0 \xi_0$.

The power spectra of the scalar and tensor perturbations were analytically
derived in (\ref{psfs}) and (\ref{psft}), respectively,
with the slow-roll approximation. Observational quantities
such as the spectral indices of the scalar and tensor modes, the tensor-to-scalar
ratio, and the running spectral indices of the scalar and tensor modes
are calculated using these power spectra to compare with
the recent CMB observation data, Planck, and BICEP2.

Regarding the tensor-to-scalar ratio, $r$, the Planck+WMAP data
provide the upper bound on $r < 0.12$, but the BICEP2 observation that
detected the B-mode polarization gives $r=0.2$. It seems
at first sight that there is some mismatch between two observations.
In Ref.\cite{Ade:2014xna}, they suggested considering the running
 spectral index to reconcile BICEP2 with  Planck.
Although there has been wide study to explain this discrepancy,
we consider in this work  the running spectral index of the scalar
perturbation by accepting the simple resolution of the
discrepancy between two data.

First, we have applied our general formalism to the large-field inflationary
model with the  exponential potential with exponential Gauss-Bonnet coupling
 (\ref{expotential}).
In the presence of the Gauss-Bonnet term,
we could find the valid model parameter ranges for inflation
to happen, unfortunately, these parameter ranges are not favored by the data.
Second, we have studied the large-field inflationary model
with the monomial potential with monomial Gauss-Bonnet coupling (\ref{powpot}).
In this scenario, $r$ is enhanced for $\alpha>0$ while
it is suppressed for $\alpha<0$.
This result is completely opposite compared to Refs. \cite{Guo:2010jr}--\cite{Jiang:2013gza},
 in which $r$ is enhanced for $\alpha<0$ and reduced for $\alpha>0$ for the model with potential,
 $V=V_{0}\phi^n$, and Gauss-Bonnet coupling, $\xi=\xi_0\phi^{-n}$.

As shown in Figs. \ref{fig_nsrn1}--\ref{fig_nsrn4},
the model parameter $\alpha$ can shift the predicted $r$ value vertically
 for the  fixed number of $e$-folds in $n_s$-$r$ plane.
For $n=1$, the theoretical predictions
can be made to better fit to  Planck data.
However, this type of model is not favored by the recent
Planck+BICEP2 data.
For $n=2$, the model with the quadratic potential can be made
a better fit to the recent Planck+BICEP2 data within a
certain parameter range given in Table \ref{tab4}.
In the model with $n=4$ for $\alpha=0$, it is well known that
this scenario of inflation is excluded by the
Planck data. However, the predictions with $\alpha \neq 0$ for $n=4$
lie inside of the $2\sigma$ contour for $N\gtrsim 60$.

However, among the model predictions in this work,
$dn_s/d\ln k$ turns out to be inconsistent with  BICEP2
combining with the Planck data  that lie outside of $2\sigma$ contour
of the BICEP2+Planck data. It would be interesting to search for
the alternatives to reconcile Planck data with BICEP2
besides consideration of the running spectral index.
We will leave it as a future work whether there are
any solutions for our model to satisfy both the BICEP2 and Planck data without
considering $dn_s/d\ln k$.

\acknowledgments

We appreciate APCTP for its hospitality during completion of this work.
We acknowledge the use of publicly available {\tt COSMOMC}.
We thank Dhiraj Kumar Hazra and Qing-Guo Huang, and Seokcheon Lee for helpful discussion.
This work was supported by the National Research Foundation of Korea(NRF) grant funded by the Korea government(MSIP)(2014R1A2A01002306). S.K was supported by Basic Science Research Program through the National Research Foundation  of Korea (NRF) funded by the Ministry of Education, Science and Technology (NRF-2010-0022596). W.L. was supported by Basic Science Research Program through the National Research Foundation of Korea(NRF) funded by the Ministry of Education, Science and Technology(2012R1A1A2043908).

\end{document}